\documentclass[rnote]{aa}
\usepackage{graphicx}
\usepackage{txfonts}

\newcommand\U[1]{{\,\rm #1}}
\newcommand\rs[1]{_\mathrm{#1}}
\newcommand\g{$\gamma$}
\newcommand\ApJ{ApJ}

\begin{document}

%%%%%%%%%%%%%%%%%%%%%%%%%%%%%%%%%%%%%%%%%%%%%%%%%%%%%%%%%%%%%%%%%%%%%%%%%
%%%%===Title=========================================================%%%%
%%%%%%%%%%%%%%%%%%%%%%%%%%%%%%%%%%%%%%%%%%%%%%%%%%%%%%%%%%%%%%%%%%%%%%%%%

\title{The artificial broadening of the high-energy end of electron spectrum in supernova remnants}
\author{O. Petruk}
%\offprints{O. Petruk}
\institute{
   Institute for Applied Problems in Mechanics and Mathematics, Naukova St.\ 3-b
   Lviv 79060, Ukraine\\
   Astronomical Observatory, National University, Kyryla and Methodia St.\ 8 Lviv 79008, Ukraine \\
   \email{petruk@astro.franko.lviv.ua}
}
\date{Received ...; accepted ...}

\abstract{}
{The observed spectrum of a supernova remnant (SNR) is a superposition of many ``local'' spectra emitted by regions of SNRs that are under different physical conditions. 
The question remains as to whether the broadening of the high-energy end of the observed nonthermal spectrum of SNRs, like in G347.3-0.5 and SN~1006, can be an artifact of observations or it is a consequence of the microphysics involved in the acceleration process.
In this note we study the influence of parameters variations (inside the volume and over the surface of SNR) on the shape of the high-energy end of the synchrotron (and also inverse Compton) spectrum.}
{We consider three possibilities for these parameter variations: i) gradients downstream of the shock with constant maximum energy of the accelerated electrons and the potential variation in time of the injection efficiency, ii) then we add the possibility of the maximum energy depending on time, and finally 
iii) the possible obliquity dependences of maximum energy and injection efficiency.}
{It is shown that gradients of density and magnetic field strength downstream of the shock are ineffective in modifying  the shape of the synchrotron spectrum, even if an SNR evolves in the nonuniform interstellar medium and/or the injection efficiency varies in time. The time dependence of the maximum energy of the electrons accelerated by the shock is also not able to make the observed spectrum much broader. The only possibility of producing considerable broadening in the spectrum is the variation in the maximum energy of electrons over the surface of SNR. In such a case, the obliquity dependence of the injection efficiency also affects the shape of the spectrum, but its role is less significant.}
{}


\titlerunning{Artificial broadening of the end of electron spectrum in SNRs}
\maketitle

%%%%%%%%%%%%%%%%%%%%%%%%%%%%%%%%%%%%%%%%%%%%%%%%%%%%%%%%%%%%%%%%%%%%%%%%%
\section{Introduction}

The properties of observed nonthermal emission from a number of supernova remnants (SNRs) in X- and \g-rays are the subject of many papers in the past decade because this emission helps for understanding a number of important questions in modern high-energy physics and astrophysics (see Jones et al. \cite{Jones-et-al-1998}; Drury et al. \cite{drury-et-al-01}; Reynolds \cite{Reynolds-2004} for reviews). 

Electrons accelerated by SNR shocks up to the maximum energies $E\rs{max}\sim 10^{13}-10^{15}\U{eV}$ radiate their energy through the synchrotron emission in X-rays and through inverse Compton or nonthermal bremstrahlung in \g-rays. 
The empirical expression for the power-law spectrum of accelerated electrons with the upper cut-off 
(e.g., Gaisser, Protheroe, \& Stanev \cite{Gaisser-Protheroe-Stanev-98})
\begin{equation}
 N(E)=KE^{-s}\exp\left[{-\left({E\over E\rs{max}}\right)^{\alpha}}\right] 
 \label{Na-fit}
\end{equation}
is robust. 
The value $\alpha=1$ produces the most rapid cutoff in comparison with any physical model considered 
(Reynolds \& Keohane \cite{Reynolds-Keoh-99} and references therein) and, 
 therefore, allows an estimate of the upper limit of $E\rs{max}$ from the X-ray spectra of SNRs 
(Reynolds \& Keohane \cite{Reynolds-Keoh-99}, Hendrick \& Reynolds \cite{Hendrick-Reynolds-2001}). 

The fit of the observed high-energy spectrum (X-ray and/or $\gamma$-ray) of the given SNR may, however, require the 
broadening of the upper cut-off to the electron spectrum; i.e. the value of $\alpha$ would be less than unity  
and the high-energy end of the emission spectrum of \object{SN 1006} is broader than could be expected in the case $\alpha=1$ (Reynolds \cite{Reyn-1996}). Detailed calculations give $\alpha\approx 0.5$ for this SNR (Ellison et al. \cite{Ell-Ber-Baring-2000}). 
The case of SNR \object{G347.3-0.5} also requires the value $\alpha\approx 0.5$ (Ellison et al. \cite{Ell-Slane-Gaensler-2001}; Uchiyama et al. \cite{uchi-ahar-2004}; Lazendic et al. \cite{Lazendic-et-al-2004}). 

Such broadening can be caused by the microphysics involved in the electron acceleration process, such as in the model of Ellison et al. (\cite{Ell-Ber-Baring-2000}). 
On the other hand, this broadening could be an artifact of observations, where the observed spectrum is coming from a certain area of the projection of SNR on the plane of the sky. There is a part of the SNR surface, as well as a part of the SNR volume, projected onto this area. Different fluid elements are under different physical conditions in different places on the surface and in the volume involved. Therefore, the spectrum observed 
is a superposition of different ``local'' spectra. This ``artificial'' emission spectrum, which is a sum of the spectra produced under different conditions, is expected to differ from the spectrum radiated by a population of electrons characterized by a single set of magneto-hydrodynamic parameters. 
In this way, the ``artificial'' broadening of the electron distribution might be a consequence of the hydro-magnetic parameters variation in space, such as the suggested variations in $E\rs{max}$ (Reynolds \cite{Reyn-1996}). Note, that we say ``broadening'' while not knowing in advance whether such an effect leads to broadening or narrowing of the spectrum end.

The measure of broadening ($\alpha$) in such a case depends on a number of factors. As an example, one of them could be  a gradient of density downstream of the shock because the shapes of nonthermal spectra are sensitive to variations in the magnetic field strength $B$ and the normalization $K$ inside the volume; both of them are functions of density $n$ whose  profile, in turn, depends on the density gradient of the interstellar medium (ISM) before the shock. 
In this note we study the influence of the nonuniformity of the ISM, as well as other factors, on 
the artificial broadening of the high-energy end of the electron spectrum in the adiabatic SNRs. 

%%%%%%%%%%%%%%%%%%%%%%%%%%%%%%%%%%%%%%%%%%%%%%%%%%%%%%%%%%%%%%%%%%%%%%%%%
\section{Models and results}
%\subsection{Synchrotron emission of uniform and nonuniform plasma}

The spectral distribution of energy radiated by a single electron with energy $E$ is
\begin{equation}
 p(E,\nu)={\sqrt{3}e^3B\rs{p}\over mc^2}F\left({\nu\over\nu\rs{c}}\right),  %erg/s/Hz
\end{equation}
where $\nu$ is frequency, $B\rs{p}=\sqrt{2/3}B$ is an angle-averaged magnetic field, $F$ and $\nu\rs{c}\propto BE^2$ are the special function and the critical frequency known from the synchrotron emission theory, $e$ and $m$ are electron charge and mass, and $c$ is the velocity of light. The synchrotron emissivity of the population of electrons is
\begin{equation}
 P(\nu)=\int_{0}^{\infty}N(E)p(E,\nu)dE  %erg/s/cm^3/Hz;    [N(E)]=cm^-3/eV, [dE]=eV
 \label{emiss}
\end{equation}
and the synchrotron spectrum from SNR is given by the integration over the given volume:
\begin{equation}
 S(\nu)=\int P(\nu)dV .
\end{equation}

\begin{figure}
   \centering
   \includegraphics[height=7.5truecm]{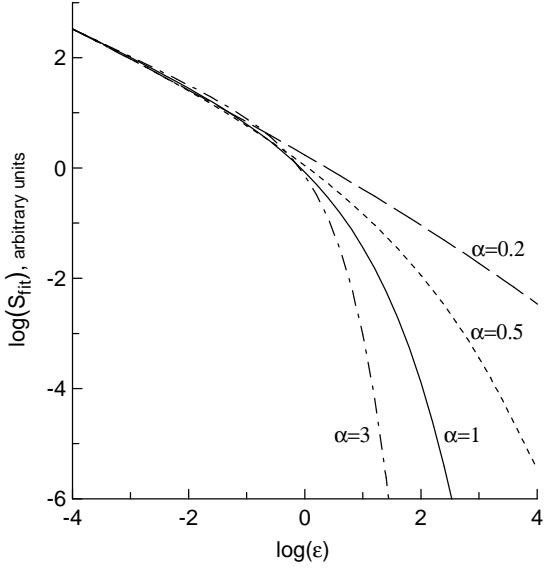}
      \caption{Influence of the broadening parameter $\alpha$ on the shape of 
       the synchrotron spectrum (\ref{sp-simpl-fit}).
      The values of $\alpha$ are marked near the respective lines, $s=2$.     
      } 
         \label{fig-a}
   \end{figure}

Uniform plasma with the electron population having a broader distribution (\ref{Na-fit}) produces a ``fit''-spectrum that does not require the volume integration
\begin{equation}
 S\rs{fit}(\nu)\propto P(\nu).
 \label{simple-spectrum-fit}
\end{equation}
It can be written in the dimensionless form
\begin{equation}
 S\rs{fit}(\varepsilon) \propto
 \int_{0}^{\infty}F\left(\varepsilon\epsilon^{-2}\right)
 \epsilon^{-2}\exp\big(-\epsilon^\alpha\big)d\epsilon,
 \label{sp-simpl-fit}
\end{equation}
where $\epsilon=E/E\rs{max}$, $\varepsilon=\nu/\nu\rs{c}(E\rs{max},B)$, and the classic $s=2$ is used. 
In these notations, most of the photons emitted by electrons with energy $\epsilon$ have energy $\varepsilon\rs{m}=0.29\varepsilon\rs{c}$, where the critical energy is $\varepsilon\rs{c}=\epsilon^2$. 
Figure~\ref{fig-a} demonstrates the role of the broadening parameter 
$\alpha$ in the modification of the shape the synchrotron spectrum.

In the following sections, we calculate the high-energy end of the synchrotron spectrum from a part of SNR assuming that $\alpha=1$ and taking into account the possible variations of parameters within the given volume (Sect.~\ref{sect-volume}) or on the given surface (Sect.~\ref{sect-surface}) of SNR in order to see whether we can obtain an artificial spectrum with a shape similar to the fit-spectrum with $\alpha=0.5$.

%--------------------------------------------------------------
%--------------------------------------------------------------

%%%%%%%%%%%%%%%%%%%%%%%%%%%%%%%%%%%%%%%%%%%%%%%%%%%%%%%%%%%%%%%%%%%%%%%%%
\subsection{Variation of the parameters inside the SNR volume}
\label{sect-volume}

Let us consider a spherical SNR on the adiabatic phase of evolution in the ISM with the power-law density variation
$\rho\rs{o}(R)=AR^{-\omega}$, where $A$ and $\omega$ are constant; indexes ``o'' and ``s'' refer to the pre- and post-shock values. The dynamics of the adiabatic shock in this medium is given by the Sedov (\cite{Sedov-59}) solutions where the shock velocity $V\propto R^{-(3-\omega)/2}$ and $R\propto t^{2/(5-\omega)}$.
These solutions are self-similar, which means that the distribution of parameters behind the shock, e.g. density $n(a,t)$, can be written as $n(a,t)=n\rs{s}(t)\bar{n}(\bar{a})$, 
where $a$ is the Lagrangian coordinate, and the overline refers to normalized parameters, i.e. ones divided on their own values at the shock front, e.g. $\bar{a}=a/R$. 

Let us assume, following most of the recent models of SNRs, that the ambient magnetic field is uniform. In the description of the magnetic-field evolution downstream we follow the model of Reynolds \& Chevalier (\cite{Reyn-Chev-1981}) and Reynolds (\cite{Reyn-98}), where  
the strength of the magnetic field is $B^2=B_{\perp}^2+B_{\parallel}^2$ with the perpendicular and parallel components evolve independently behind the shock:  
$\bar{B}_{\perp}=\bar{n}{\bar{r}}/{\bar{a}},$ $\bar{B}_{\parallel}=\left({\bar{a}}/{\bar{r}}\right)^2$. 
In this case, the behaviour of the magnetic field is also self-similar: 
$B(a,\Theta\rs{o})=B\rs{o}\sigma\rs{B}(\Theta\rs{o})\bar B(\bar a,\Theta\rs{o})$ where the magnetic field compression factor $\sigma\rs{B}=\left(\left({1+16 \tan^2\Theta\rs{o}}\right)/\left({1+\tan^2\Theta\rs{o}}\right)\right)^{1/2}$.

We also follow Reynolds (\cite{Reyn-98}) in the description of the evolution of electron distribution. 
Fluid element $a\equiv R(t\rs{i})$ was shocked at time $t\rs{i}$. At that time 
the electron distribution on the shock was 
\begin{equation}
 N(E\rs{i},t\rs{i})=K\rs{s}(t\rs{i})E\rs{i}^{-s}
 \exp\left[{-\left({E\rs{i}\over E\rs{max}}\right)}\right] 
 \label{sp-ini}
\end{equation}
(the value $s=2$ is used throughout this paper). 

An electron loses its energy downstream due to the adiabatic expansion and radiative losses 
(synchrotron and inverse Compton on a cosmic microwave background). 
Its energy varies as (Reynolds \cite{Reyn-98})
\begin{equation}
 E=E\rs{i}\bar{n}(\bar{a})^{1/3} \big(1+{\cal I}(\bar{a}){E\rs{i}}/{E\rs{f}}\big)^{-1}
 \label{energylosses}
\end{equation} 
where ${\cal I}$ is the dimensionless function
\begin{equation}
 {\cal I}(\bar{a},\Theta\rs{o},\omega,d)=
 \frac{5-\omega}{2}\int_{\bar{a}}^{1} x^{(3-\omega)/2}{\bar{B}\rs{eff}\left({\bar{a}\over x}\right)^2}
 \bar{n}\left({\bar{a}\over x}\right)^{1/3}dx,
 \label{Int}
\end{equation} 
$d=B\rs{CMB}/B\rs{o}$, $B\rs{CMB}=3.27\U{\mu G}$ is the magnetic field strength with energy density equal to that in CMB, and 
\begin{equation}
 E\rs{f}=637/(B\rs{eff,s}^{2}t)\U{erg}=13B_5^{-2}t_4^{-1}\U{TeV}
 \label{fiduc-energy}
\end{equation}
is the fiducial energy, ${B}\rs{eff}^2={B}^2+{B}\rs{CMB}^2$, $B_5=10^5B\rs{eff,s}$, $t_4=10^{-4}t$. 

The conservation law for the number of particles per unit volume per unit energy interval 
\begin{equation}
 N(E,a,t)=N(E\rs{i},a,t\rs{i}){a^2dadE\rs{i}\over 4r^2drdE}, 
 \label{cons-N}
\end{equation}
together with the continuity equation $\rho_{\rm o}(a)a^2da=\rho(a,t)r^2dr$ 
and the derivative $dE\rs{i}/dE=\bar n^{1/3}{E\rs{i}^2}/{E^2}$, 
implies that downstream 
\begin{equation}
 N(E,a,t)=KE^{-2}
  \exp\left[{-\left({E\rs{i}(E)\over E\rs{max}}\right)}\right]
\label{N-downstr}
\end{equation}
with
\begin{equation}
 K(a,t)=K_{\rm s}(t\rs{i})\ \bar{n}^{4/3}\ \bar{a}^{\ \!\omega}.
\end{equation}
If $K_{\rm s}\propto V(t)^{-b}$, the evolution of $K$ is also self-similar downstream:
\begin{equation}
 \bar{K}(a)=K(a,t)/K_{\rm s}(t)=
 \bar{a}^{\ \!(3b+\omega(2-b))/2}\ 
 \bar{n}(\bar{a})^{4/3}.
 \label{self-K}
\end{equation}
It is unknown how the injection efficiency (on which $K$ depends) evolves in time. Reynolds (\cite{Reyn-98}) considered three typical alternatives for $b$ as a free parameter, namely, $b=0,-1,-2$ (e.g. if acceleration efficiency is a constant fraction of postshock pressure, then $K\propto V^2$). Petruk \& Bandiera (\cite{pet-band-2006}) show that one can expect $b>0$ and its value can even be as high as $b\approx 5$. 

The dependence $E\rs{i}(E)$ used in (\ref{N-downstr}) can be obtained trivially from Eq.~(\ref{energylosses}). It may be written in the form 
$E\rs{i}/E={\cal E}(\bar{a},E)^{-1}$, where 
\begin{equation}
 {\cal E}=\bar{n}(\bar{a})^{1/3}-{\cal I}(\bar{a},\Theta\rs{o})E/E\rs{f}.
\end{equation} 

%--------------------------------------------------------------
   \begin{figure*}
   \centering
   \includegraphics[height=7.5truecm]{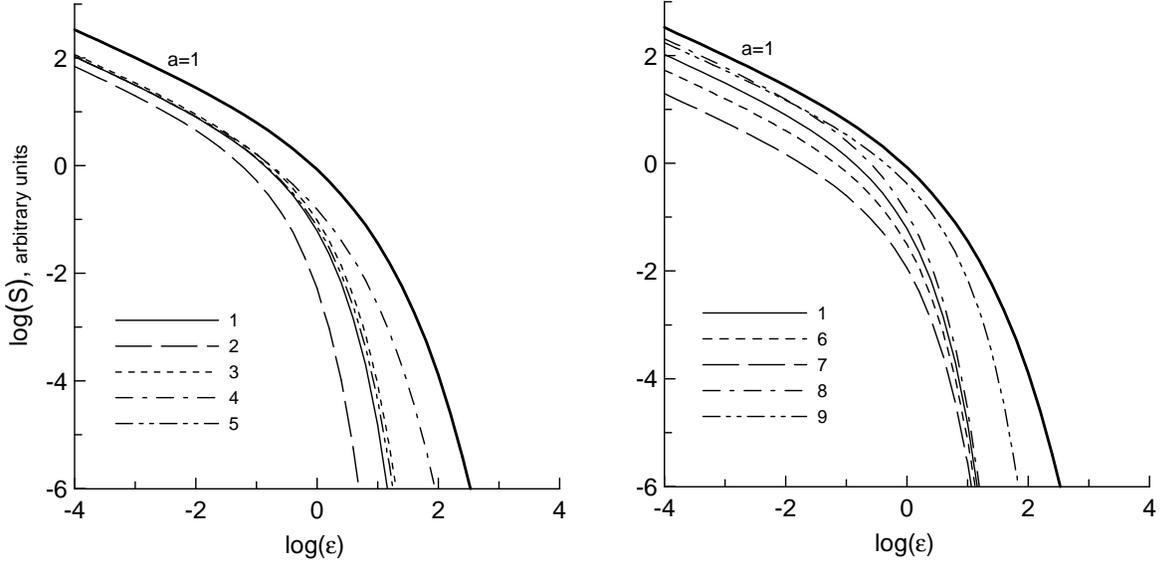}
      \caption{Evolution of the synchrotron spectrum downstream of the shock.
      The thick solid line shows spectra for different sets of parameters at $\bar a=1$. 
      Other lines correspond to spectra at $\bar a=0.8$. 
      The basic set of parameters is $\Theta\rs{o}=0$, $\omega=0$, $d=1$, $\epsilon\rs{f}=1$, $b=0$. 
      The corresponding spectrum is shown by line 1. Other lines show spectra with only one  
      parameter changed: 
      2 -- $\omega=-5$,
      3 -- $\omega=1$,
      4 -- $\epsilon\rs{f}=10$,
      5 -- $d=0.1$,
      6 -- $b=2$,
      7 -- $b=5$,
      8 -- $b=-2$,
      9 -- $\Theta\rs{o}=\pi/2$.
      } 
         \label{fig-b}
   \end{figure*}
%--------------------------------------------------------------

In order to see the effect of the superposition of spectra from different regions inside the volume of SNR on the shape of the summary spectrum, we perform an integration along the radius of a spherical SNR 
\begin{equation}
 S(\nu)%=\int_{0}^{2\pi}d\phi\int_{0}^{\pi}\sin\theta d\theta
 \propto\int_{{r}\rs{min}}^{R}P(\nu){r}^2d{r}\ . %erg/s/Hz 
 \label{simple-spectrum}
\end{equation}
Note that here we also take into account the geometrical dilution effect that takes place in spherical shocks (in contrast to planar shocks). The highest energy particles were injected when the shock was young and, therefore, had a much smaller surface area than at the observation time. Potentially different injection rates at earlier times are involved through parameter $b$.

Spectrum (\ref{simple-spectrum}) may be written in a dimensionless form for the Sedov SNR:
\begin{equation}
\begin{array}{l}
 \displaystyle
 S(\varepsilon) 
 \propto\int_{\bar{a}\rs{min}}^{1}d\bar{a}\ \bar{r}\rs{\bar{a}}(\bar{a})\bar{r}(\bar{a})^2
  \bar{B}(\bar{a},\Theta\rs{o})\bar{K}(\bar{a})
 \\ \\ \displaystyle\qquad
 \times\int_{0}^{\infty}F\left(\frac{\varepsilon}{\epsilon^{2}\sigma\rs{B}\bar B}\right)
 \epsilon^{-2}\exp\left[-\frac{\epsilon}{
{\cal E}(\bar{a},\Theta\rs{o},\omega,d,\epsilon/\epsilon\rs{f})
 }\right]d\epsilon
\end{array} 
 \label{sp-simpl}
\end{equation}
where $\bar{r}\rs{\bar{a}}=d\bar{r}/d\bar{a}$ and $\varepsilon=\nu/\nu\rs{c}(E\rs{max},B\rs{o})$. Note that, contrary to the case of uniform plasma, the electrons with energy $\epsilon$ now have critical energy $\varepsilon\rs{c}=\epsilon^2\sigma\rs{B}\bar B$.

Spectrum (\ref{sp-simpl}) depends on a set of parameters, $\Theta\rs{o}$, $\omega$, 
$d$ (i.e. on $B\rs{o}$), $\epsilon\rs{f}$ (i.e. on the combination $B\rs{eff,s}E\rs{max}t$) and $b$, while 
the ``fit''-spectrum (\ref{sp-simpl-fit}) is only a function of $\alpha$. 
It is clear from Eq.~(\ref{sp-simpl}) that the shape of the spectrum emitted by the electron population immediately postshock 
(i.e. at $\bar a=1$) is exactly the same for any set of these parameters, because 
$\bar B$, $\bar K$, and ${\cal E}$ equal unity immediately after the shock. The shape of this spectrum coincides with the shape of the ``fit''-spectrum (\ref{sp-simpl-fit}) with $\alpha=1$. The downstream evolution of the postshock spectrum depends on a set of parameters. 

Examples of the spectra at $\bar a=0.8$ are shown in Fig.~\ref{fig-b}. 
The function ${\cal E}$ is almost independent of $d$ and obliquity, if considered close to the shock (up to $\bar a\approx 0.8$). Therefore, spectra for different $d$ and $\Theta\rs{o}$ are quite close. In agreement with Reynolds (\cite{Reyn-98}), the maximum losses are on maximum electron energies. Thus, the end of the spectrum becomes steeper with time (with the decrease in $\bar a$). 
Lines 1, 5, 6, 7 on Fig.~\ref{fig-b} show spectra for different laws of evolution of $K$. It is apparent that no value of $b$ produces the broadening of the spectrum. The only possibility of reducing losses at high energies is the high  value of $\epsilon\rs{f}\propto (B\rs{eff,s}E\rs{max}t)^{-1}$ (line 4). 
Lines 1, 2, and 3 demonstrate the role of ISM nonuniformity. The shape of the spectrum is almost insensitive to $\omega$ while the amplitude is sensitive. If the ISM density increases ($\omega<0$), then the amplitude of the spectrum is less. If the density decreases, then plasma produces more emission at the same $\bar a$.  
The stronger gradient of the post-shock density for smaller $\omega$ is the reason for this behaviour. 

\begin{figure}
   \centering
   \includegraphics[height=7.5truecm]{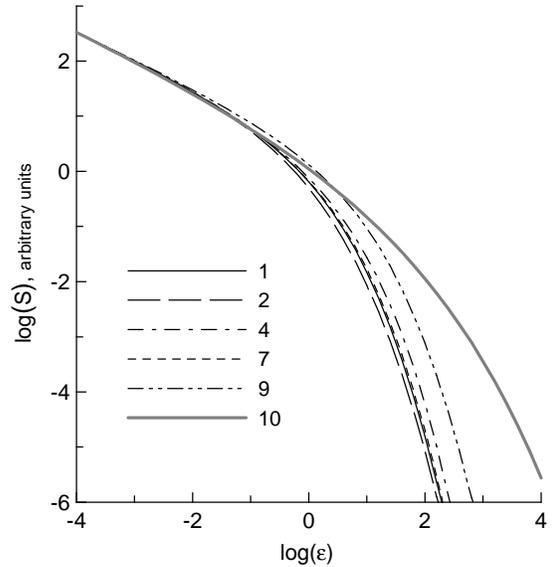}
      \caption{High-energy end of the synchrotron spectrum (\ref{sp-simpl}), $\bar a\rs{min}=0.9$.
      Lines correspond to models listed on Fig.~\ref{fig-b}. 
      All the presented spectra are normalized at $\varepsilon=10^{-4}$. 
%      All of them lie in between fit-spectra with $\alpha=0.8$ and $\alpha=1.3$.
      The fit-spectrum (\ref{sp-simpl-fit}) coinsides with line 1, if $\alpha=1$, and 
      is shown by line 10 for $\alpha=0.5$. 
      } 
         \label{fig-c}
   \end{figure}

The spectra calculated with Eq.~(\ref{sp-simpl}) are shown in Fig.~\ref{fig-c}. 
The higher the electron energy, the greater the radiative losses. 
Therefore, the  superposition of spectra from different regions inside the volume of SNR might a priori be expected to be narrower than the initial postshock spectrum. For example, line 1 may be approached by the fit-spectrum with $\alpha=1.1$; i.e. it is steeper than the initial post-shock spectrum at $\bar a=1$. All the spectra presented in Fig.~\ref{fig-c} lie between fit-spectra with $\alpha=0.8$  and $\alpha=1.3$. 
Note that Fig.~\ref{fig-c} shows spectra for those sets of input parameters that produce maximum differences in spectra at  $\bar a=0.8$ (Fig.~\ref{fig-b}). None of these sets produces a broad spectrum (\ref{sp-simpl}), which could be approximated by $S\rs{fit}$ with $\alpha\approx 0.5$. The single case of the perpendicular shock (line 9) produces a bit broader emission spectrum (close to $S\rs{fit}$ with $\alpha\approx0.8$) in comparison with the parallel-shock case. The reason for such behaviour is that electrons with the same energy $\epsilon$ are responsible for the synchrotron emission with higher energy ($\varepsilon\rs{c}=\epsilon^2\sigma\rs{B}\bar B$) in the perpendicular shock ($\sigma\rs{B}=4$ for $\Theta\rs{o}=\pi/2$ while $\sigma\rs{B}=1$ for $\Theta\rs{o}=0$). 

The efficiency of the space parameter variation in the modification of the shape of the high-energy end of the synchrotron spectrum is low. For example, line 4 differs from the line 1 maximum $1.25$ times. 
The reason for this is a rapid fall in the amplitude of spectra with the decrease in $\bar a$ caused by synchrotron losses and hydrodynamic properties. Therefore, most of the contribution to the integral (\ref{sp-simpl}) is given by the regions quite close to the shock, with $\bar a\approx 1$. 

%%%%%%%%%%%%%%%%%%%%%%%%%%%%%%%%%%%%%%%%%%%%%%%%%%%%%%%%%%%%%%%%%%%%%%%%%
\subsection{Time evolution of $E\rs{max}$}

The model in Sect.~\ref{sect-volume} assumes that the shock accelerates electrons to the same maximum energy $E\rs{max}$ at any time. It could be expected that the shock -- during its evolution -- is able to accelerate particles to different maximum energies. Reynolds (\cite{Reyn-98}) considered models where $E\rs{max}\propto V^q$ with $q=1$ if the maximum energy is limited by radiative losses, or $q=2$ if $E\rs{max}$ is determined by the finite time of acceleration. 

If the maximum energy varies in time as $E\rs{max}\propto V(t)^q$, then the spectrum (\ref{simple-spectrum}) may be calculated as 
\begin{equation}
\begin{array}{l}
 \displaystyle
 S(\varepsilon) 
 \propto\int_{\bar{a}\rs{min}}^{1}d\bar{a}\ \bar{r}\rs{\bar{a}}(\bar{a})\bar{r}(\bar{a})^2 
 \bar{B}(\bar{a},\Theta\rs{o})\bar{K}(\bar{a})
 \\ \\ \displaystyle\qquad
 \times\int_{0}^{\infty}F\left(\frac{\varepsilon}{\epsilon^{2}\sigma\rs{B}\bar B}\right)
 \epsilon^{-2}\exp\left[-\frac{\epsilon\ \bar a^{(3-\omega)q/2}}{
 {\cal E}(\bar{a},\Theta\rs{o},\omega,d,\epsilon/\epsilon\rs{f})}
 \right]d\epsilon\ .
\end{array} 
 \label{sp-simpl-Et}
\end{equation}
Here we redefine $\epsilon=E/E\rs{max}(t)$ and $\varepsilon=\nu/\nu\rs{c}(E\rs{max}(t),B\rs{o})$, where $t$ is the time of observations. 

In media with $\omega<3$, positive $q$ broadens spectra for different $\bar a$ compared with the case $q=0$. The larger $q$, the broader the specrum for the given $\bar a$. However, this broadening is too small to have a prominent effect on the integral spectrum (\ref{sp-simpl-Et}). 
The time evolution of $E\rs{max}$ is inefficient in producing broader artificial spectra. Namely, 
for the basic set of parameters (see caption to Fig.~\ref{fig-b}), the spectrum (\ref{sp-simpl-Et}) with $q=1$ ($q=2$), normalized at $\varepsilon=10^{-4}$, differs from the spectrum with $q=0$ maximum 1.15 (1.35) times. 

The reason for this inefficiency lies again in the rapid decrease in $B$ and $K$ downstream of the shock that causes small weight in the integral (\ref{sp-simpl-Et}) of the downstream layers with broader spectra in comparison with the layers $\bar a\approx1$. 

%--------------------------------------------------------------
%--------------------------------------------------------------

%%%%%%%%%%%%%%%%%%%%%%%%%%%%%%%%%%%%%%%%%%%%%%%%%%%%%%%%%%%%%%%%%%%%%%%%%
\subsection{Variation in parameter on the surface of the SNR}
\label{sect-surface}

In this section we consider neither the downstream evolution of the initial spectrum (\ref{sp-ini}) generated by a shock nor the distribution of parameters downstream. The role of the parameter variation inside the volume of SNR in the modification of the artificial spectrum is minor, as shown above. The purpose of the present section is to study a possible variation in the parameters on the shell of SNR. Three of them can influence the synchrotron spectrum: $\vec B$, $K$, $E\rs{max}$. Their changes on the shock can be caused by non uniformity of the ambient magnetic field ($B$) and/or obliquity ($K$, $E\rs{max}$). In this paper we are considering the only uniform magnetic field and spherical SNRs, hence, the nonuniformity of ISM where the density changes along the surface of the SNR is not studied here (it could be one more reason for the variation in $K$ (which depends on density) and $E\rs{max}$ (which depends on shock velocity)). 

Let $\theta$ be  an angle between the direction of $\vec B\rs{o}$ and the given direction. The variations in parameters are given by $E\rs{max}(\theta)=E\rs{*}f\rs{E}(\theta)$, $K\rs{s}(\theta)\propto f\rs{K}(\theta)$, $B\rs{s}(\theta)\propto  \sigma\rs{B}(\theta)$, where $E\rs{*}$ is a constant. 
Integration of the emissivity (\ref{emiss}) over the surface of SNR between $\theta=0\div\theta\rs{m}$ (a polar cap)  instead of (\ref{sp-simpl}) gives:  
\begin{equation}
\begin{array}{l}
 \displaystyle
 S(\varepsilon) 
 \propto\int_{0}^{\theta\rs{m}}d\theta  
 \frac{\sigma\rs{B}(\theta)f\rs{B}(\theta)f\rs{K}(\theta) \tan\theta}
 {\cos^2\theta\sqrt{1-\cos^2\theta\rs{m}\tan^2\theta}}
 \\ \\ \displaystyle\qquad
 \times\int_{0}^{\infty}F\left(\frac{\varepsilon}{\epsilon^{2}\sigma\rs{B}(\theta)}\right)
 \epsilon^{-2}\exp\left[-\frac{\epsilon}{f\rs{E}(\theta)}\right]d\epsilon,
\end{array} 
 \label{sp-simpl-surf}
\end{equation}
where $\epsilon=E/E\rs{*}$, $\varepsilon=\nu/\nu\rs{c}(E\rs{*},B\rs{o})$. 
We adopt
\begin{equation}
 f\rs{E}(\theta)=\exp\left(a\rs{E}\big({\theta}/{\theta\rs{E}}\big)^2\right)
\end{equation}
\begin{equation}
 f\rs{K}(\theta)=\exp\left(-\big({\theta}/{\theta\rs{K}}\big)^2\right),
 \label{fK}
\end{equation}
where $\theta\rs{E}$ and $\theta\rs{K}$ are some constants and $a\rs{E}$ is either $1$ or $-1$. 
Expression (\ref{fK}) is able to approximately restore the dependence of the injection efficiency on obliquity from Ellison et al. (\cite{ell-bar-jones-95}) with $\theta\rs{K}=\pi/9\div \pi/4$.

\begin{figure}
   \centering
   \includegraphics[height=7.5truecm]{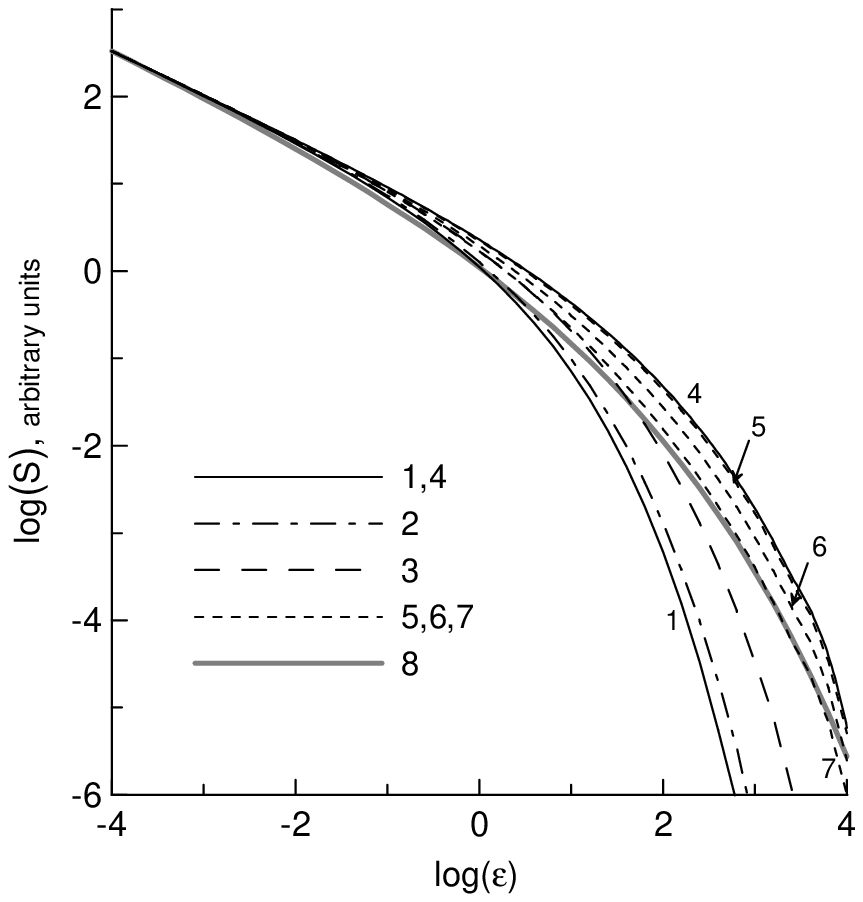}
      \caption{High-energy end of the synchrotron spectrum (\ref{sp-simpl-surf}), $\theta\rs{m}=\pi/6$, $a\rs{E}=1$. 
      Lines: 
      1 -- $\theta\rs{K}=\infty$, $\theta\rs{E}=\infty$;
      2 -- $\theta\rs{K}=\infty$, $\theta\rs{E}=\pi/3$;
      3 -- $\theta\rs{K}=\infty$, $\theta\rs{E}=\pi/6$;
      4 -- $\theta\rs{K}=\infty$, $\theta\rs{E}=\pi/9$;
      5 -- $\theta\rs{K}=\pi/4$, $\theta\rs{E}=\pi/9$;
      6 -- $\theta\rs{K}=\pi/9$, $\theta\rs{E}=\pi/9$;
      7 -- $\theta\rs{K}=\pi/12$, $\theta\rs{E}=\pi/9$.
      All presented spectra are normalized at $\varepsilon=10^{-4}$. 
      The fit-spectrum (\ref{sp-simpl-fit}) with $\alpha=0.5$
      is shown by line 8. 
      } 
         \label{fig-d}
   \end{figure}

Equation~(\ref{sp-simpl-surf}) shows that the shape of the high-energy end of the emission spectrum does not depend on $f\rs{K}(\theta)$ if $f\rs{E}(\theta)=\mathrm{const}$ (in this case the spectrum shape coincides with the shape of  fit-spectrum $S\rs{fit}$ with $\alpha=1$). The gradients in the distribution of $E\rs{max}$ plays the main role in the modification of the shape of spectrum (\ref{sp-simpl-surf}). If $a\rs{E}=1$, then one has a broader spectrum (Fig.~\ref{fig-d}, lines 1-4), and a narrower one if $a\rs{E}=-1$. If $E\rs{max}$ varies, then changes in $K$ also affect on the shape of the spectrum (Fig.~\ref{fig-d}, lines 4-7), but its role is less efficient. 

%--------------------------------------------------------------
%--------------------------------------------------------------

%%%%%%%%%%%%%%%%%%%%%%%%%%%%%%%%%%%%%%%%%%%%%%%%%%%%%%%%%%%%%%%%%%%%%%%%%
\section{Conclusions}

1. The internal property of the electron spectrum produced on the shock can be one of the reasons for the observed broadening of the high-energy end of the synchrotron and inverse Compton spectra of SN~1006 and G347.3-0.5. This broadening consists in $\alpha<1$ and is an intrinsic property of processes involved in the formation of the spectrum in the region of acceleration, as in a model of Ellison et al. (\cite{Ell-Ber-Baring-2000}). 

2. The broadening of the observed spectrum could be an artifact of observations. This means that electrons were accelerated on the shock and their real $\alpha=1$, but the observed synchrotron spectrum looks as it is produced by electrons with $\alpha<1$. This could be caused by the fact that the observed spectrum is a superposition of spectra from different regions of SNR. However, both the variation in parameters downstream of the shock (even strong variation if SNR is in a nouniform ISM) and the time evolution of the injection efficiency and/or $E\rs{max}$ are ineffective in the production of a broader artificial spectrum. The shape of this a spectrum is in general narrower than the shape of the spectrum emitted by the electron population immediately after the shock. Such an artificial spectrum can be approximated by $S\rs{fit}$ with $\alpha\approx 1\div 1.3$ for different sets of parameters. Only perpendicular shocks are able to put the observed synchrotron spectrum close to $S\rs{fit}$ with $\alpha=0.8$ due to a higher compression of the magnetic field that makes the emitted photons harder (the critical frequency $\varepsilon\rs{c}\propto \sigma\rs{B}(\Theta\rs{o})$).

3. If the broadening of the spectrum is due to the variation of parameters, there has to be a variation in $E\rs{max}$ on the surface of SNR. The gradients $dE\rs{max}/d\theta$ have to be fairly strong to produce spectra close to $S\rs{fit}$ with $\alpha=0.5$. Namely, if the spectrum is observed from a region of SNR, like a polar cap between $\theta =0\div \theta\rs{m}$, then $\theta\rs{E}\approx 2\theta\rs{m}/3$ and, therefore, $E\rs{max}$ has to change inside this cap almost $10$ times. 

%%%%%%%%%%%%%%%%%%%%%%%%%%%%%%%%%%%%%%%%%%%%%%%%%%%%%%%%%%%%%%%%%%%%%%%%%

\end{document}